\documentclass[12pt,twoside]{article}
\usepackage[dvips]{graphicx}
\usepackage{float}
\usepackage{latexsym}
\usepackage{amsmath,mathrsfs,bm,amssymb,color}

\newtheorem{theorem}{Theorem}[section]
\newtheorem{proposition}[theorem]{Proposition}
\newtheorem{lemma}[theorem]{Lemma}
\newtheorem{corollary}[theorem]{Corollary}
\newtheorem{definition}[theorem]{Definition}
\newtheorem{example}[theorem]{Example}
\newtheorem{remark}[theorem]{Remark}

\newtheorem{assumption}[theorem]{Assumption}
\newcommand{\QQ}{{L^2(Q)}}
\newcommand{\QQQ}{{L^2(Q_E)}}
\renewcommand{\AA}{\ms A_E}

\newcommand{\ms}{\mathscr}
\newcommand{\jjj}{\sum_{j=1}^{d-1}}
\newcommand{\mmm}{\sum_{\mu=1}^d}
\newcommand{\jj}{{\rm j}}
\newcommand{\JJ}{{\rm J}}

\newcommand{\dV}{\,\dot +\,V_+\, \dot -\, V_-}

\newcommand{\eq}[1]{\begin{equation}\label{#1}}
\newcommand{\en}{\end{equation}}
\newcommand{\pro}[1]{(#1_t)_{t\geq 0}}

\newcommand{\eqn}{
\begin{eqnarray*}}
  \newcommand{\enn}{\end{eqnarray*}}
\newcommand{\proof}{{\noindent \it Proof:\ }}
\newcommand{\qed}{\hfill ${\rm QED}$\par\medskip}
\newcommand{\BR}{{{\mathbb  R}^d}}
\newcommand{\BRT}{{{\mathbb  R}^{d+1}}}
\newcommand{\PF}{H_{\rm PF}}
\newcommand{\PPF}{\widehat H_{\rm PF}}

\newcommand{\cmp}[5]{{#1}, {#2}, {\it Commun. Math. Phys.} {\bf #3} (#4), #5}

\newcommand{\bi}{\begin{description}}
\newcommand{\ei}{\end{description} }
\newcommand{\CC}{{{\mathbb  C}}}
\newcommand{\RR}{{\mathbb  R}}
\newcommand{\hp}{H_{\rm p}}
\newcommand{\bl}[1]{\begin{lemma}\label{#1}}
\newcommand{\el}{\end{lemma}}
\newcommand{\bc}[1]{\begin{corollary}\label{#1}}
\newcommand{\ec}{\end{corollary}}
\newcommand{\bt}[1]{\begin{theorem}\label{#1}}
\newcommand{\et}{\end{theorem}}
\newcommand{\bp}[1]{\begin{proposition}\label{#1}}
\newcommand{\ep}{\end{proposition}}
\newcommand{\br}[1]{\begin{remark}\label{#1}}
\newcommand{\er}{\end{remark}}

\newcommand{\sr}{\sqrt\alpha}

\newcommand{\kak}[1]{(\ref{#1})}

\newcommand{\LR}{{L^2(\BR)}}
\newcommand{\LRx}{{L^2(\RR^d_x)}}
\newcommand{\LRk}{{L^2(\RR^d_k)}}
\newcommand{\LRT}{{L^2(\BRT)}}
\newcommand{\LRR}{{L^\infty(\RR^d_x;L^2(\RR^d_k))}}

\newcommand{\cb}[1]{C_{\rm b}^{#1}(\RR^d_x;L^2(\RR^d_k))}

\newcommand{\fff}{{\cal F}}
\newcommand{\is}{\inf\!\sigma}

\newcommand{\f}{^{-1}}
\newcommand{\lk}{\left(}
\newcommand{\rk}{\right)}
\newcommand{\lkk}{\left\{}
\newcommand{\rkk}{\right\}}
\newcommand{\lkkk}{\left[}
\newcommand{\rkkk}{\right]}
\newcommand{\Ebb}{\mathbb E}

\newcommand{\add}{a^{\ast}}

\newcommand{\ov}[1]{\overline{#1}}
\newcommand{\hf}{H_{\rm f}(m)}
\newcommand{\gr}{\varphi_{\rm b}}

\newcommand{\half}{\frac{1}{2}}
\newcommand{\han}{{1/2}}

\newcommand{\la}{\lambda}

\newcommand{\hhh}{{\ms H}}

\renewcommand{\d}{\displaystyle}
\newcommand{\non}{\nonumber}
\newcommand{\od}{\oplus^{d-1}}
\newcommand{\V}{\int_0^tV(B_s) ds}
\newcommand{\odj}{\oplus_{j=1}^{d-1}}

\newcommand{\KPF}{K_{\rm PF}}
\newcommand{\loc}{L_{\rm loc}^1(\BR)}

\usepackage{amsmath,amssymb}

\usepackage{graphicx,color}

\makeatletter
\@addtoreset{equation}{section}
\makeatother

\title
{\sc Pauli-Fierz model with Kato-class potentials and exponential decays}

\author{\small
Takeru Hidaka
\thanks{
Faculty of Mathematics,
Kyushu University, Fukuoka 819-0385, Japan.}
 and  Fumio Hiroshima
\thanks{
Faculty of Mathematics,
Kyushu University, Fukuoka 819-0385, Japan.}
}\date{\today}
\begin{document}
\pagestyle{myheadings}
\markboth{Generalized PF model}
{Generalized PF model}

\setlength{\baselineskip}{20pt} \maketitle

\begin{abstract}
Generalized Pauli-Fierz Hamiltonian with Kato-class potential $\KPF$ in nonrelativistic quantum electrodynamics is defined and studied by a path measure.
$\KPF$ is defined as the self-adjoint generator
of a strongly continuous one-parameter symmetric semigroup
and
it is shown that
its bound states
spatially exponentially decay pointwise and the ground state is unique.
\end{abstract}

\section{Introduction}
In this paper we investigate
generalized Pauli-Fierz Hamiltonians with Kato-class  potentials in nonrelativistic quantum electrodynamics by a path measure.
It includes not only Kato-class potentials but also general cutoff functions of quantized radiation fields. Basic ingredients in this paper are path measures  and functional integral representation of semigroups.
It has been shown that
 functional integral representations are useful tools to investigate the spectrum of models in quantum field theory. See  e.g., \cite{bhlms,gub, h9, h26,hl, lms1,nel, sp3,sp4}.

The strongly continuous one-parameter semigroup $(e^{-t\hp})_{t\geq 0}$  generated by
the Schr\"odinger operator,
$\hp=\half (p-a)^2+V$,
on $\LR$
with some
external potential $V$ and vector potential $a=(a_1,\cdots,a_d)$
is expressed by a path measure,  which is known as
 Feynman-Kac-It\^o formula \cite{sim79}:
\eq{m12}
(f, e^{-t\hp} g)=\int dx\bar f(x)
\Ebb^x\lkkk
 e^{-\int_0^t V(B_s) ds-i\int_0^t a(B_s)\circ dB_s}g(B_t)\rkkk,
 \en
 where $\Ebb^x$ denotes the expectation value with respect to the Wiener measure $P^x$,  $\pro B$ the $d$-dimensional Brownian motion and $\int _0^t a(B_s)\circ dB_s$ a Stratonovich integral.

Conversely since  a Kato-class potential $V$ satisfies that
\eq{to1}
\sup_x\Ebb^x\lkkk
 e^{-\int_0^t V(B_s) ds}\rkkk<\infty,\quad  t\geq0,
 \en
 the family of mappings $S_t$ defined by
  \eq{ko1}
S_tg(x)=\Ebb^x\lkkk
 e^{-\int_0^t V(B_s) ds-i\int_0^t a(B_s)\circ dB_s}g(B_t)\rkkk,\quad t\geq0,\en
turns to be the strongly continuous one-parameter symmetric semigroup
for a Kato-class potential $V$.
The Schr\"odinger operator
with
a Kato-class  potential $V$
is then defined as the self-adjoint generator of
$\pro S$.
See e.g., \cite{bhl,sim82,hil}.
Three-dimensional Kato class  includes a singular external potential such as $V(x)=-|x|^{-a}$, $0\leq a<2$.

We extend this to the Pauli-Fierz Hamiltonian.
The Pauli-Fierz Hamiltonian $\PF$ is a self-adjoint operator
defined on the tensor product of Hilbert spaces:
 \eq{ta1}
\hhh=\LR\otimes\QQ,\en
where $\QQ$ is an $L^2$-space over a probability apace
$(Q, \ms B, \mu)$ with a Gaussian measure $\mu$, and
it describes the Schr\"odinger representation of the standard Boson Fock space.
The Pauli-Fierz Hamiltonian $\PF$ is
given by
\eq{m1}
\PF=\half(p\otimes 1+\sqrt\alpha \ms A)^2+V\otimes 1 +1\otimes \hf,
\en
where $\alpha\geq 0$ is a coupling constant,  $\hf$  the free field Hamiltonian with a field mass $m\geq0$
and $\ms A=(\ms A_1,...,\ms A_d)$  a quantized radiation field with a cutoff function.
See Section 2 for the detail of  notations.
Under some conditions on cutoff functions  and $V$ it is proven that
\kak{m1} is self-adjoint and $e^{-t\PF}$ is then defined by the spectral resolution.
In  \cite{h4},
$(F, e^{-t\PF} G)$ is
also presented
by a path measure:
\eq{m11}
(F, e^{-t\PF} G)=\int dx \lk
F(x), (T_tG)(x)\rk_\QQ,
\en
where
$T_t$ is of the form \eq{na2}
T_tf(x)=\Ebb^x\lkkk
e^{-\int_0^t V(B_s) ds}
\JJ_0^\ast e^{i\sqrt\alpha\AA(K_t)}
\JJ_t  G(B_t)\rkkk\in\QQ
\en for each $x\in\BR$.
Compare with \kak{ko1} and see \kak{5} for the detail.

Our construction of generalized Pauli-Fierz Hamiltonians is closed to the procedure to define the Schr\"odinger operator with Kato-class potentials. We believe however that it is worthwhile extending  it to the Pauli-Fierz Hamiltonian from mathematical point of view.
It will be shown that
the family of operators $T_t:\hhh\to\hhh$, $t\geq0$,  can be also defined
for Kato class potentials $V$ and general cutoff functions in $\ms A$,
and the generalized  Pauli-Fierz Hamiltonian
$\KPF$ is defined as the self-adjoint generator of $\pro T$.
Of course under some  conditions $\KPF$ coincides with $\PF$, but $\KPF$ permits  to include more singular  V's and general cutoff functions in $\ms A$.

Cutoff functions of $\ms A_\mu(x)$, $\mu=1,2,3$, of the  standard Pauli-Fierz Hamiltonian in three-dimension are of the form
\eq{ka4}
e^{-ikx}
e_\mu(k,j)\hat \varphi(k)/\sqrt{|k|}
\en
 with some function $\hat \varphi$ and polarization vectors $e(k,j)=(
 e_1(k,j),e_2(k,j),e_3(k,j))$, $j=1,2$.
In \cite{ghps} the so called Nelson model
on a pseudo Riemannian manifold is studied by a path measure.
Generalized Pauli-Fierz Hamiltonians  include a mathematical analogue of the Nelson model on a pseudo Riemannian manifold, which is unitarily transformed to the Pauli-Fierz Hamiltonian with a variable mass.
Cutoff function of
the Pauli-Fierz Hamiltonian with a variable mass $v$ is \kak{ka4}
  with $e^{ikx}$ and $e_\mu(k,j)\hat \varphi(k)$
replaced by $\Psi(k,x)$ and $\hat\phi_\mu^j(k)$, respectively:
\eq{ka2}
\ov{\Psi(k,x)}
\hat\phi_\mu^j(k)
/\sqrt{|k|}.
\en
Here $\hat\phi_\mu^j(k)
$ is some function and $\Psi(k,x)$, $k\not=0$,  is the unique solution of
the Lippman-Schwinger equation \cite{ike}:
\eq{lipp0}
\Psi(k,x)=e^{+ikx}-\frac{1}{4\pi}\int\frac{e^{i|k||x-y|}
v(y)}{|x-y|}\Psi(k,y) dy.
\en
The main results of the present paper are as follows:
\bi
\item[(1)]
we define the generalized Pauli-Fierz Hamiltonian $\KPF$ with Kato class potentials and generalized cutoff functions, i.e., we prove that $\pro T$ is a strongly continuous one-parameter symmetric semigroup;
\item[(2)]
$\KPF$ is an extension of $\PF$;
\item[(3)]
bound states of $\KPF$ spatially  exponentially decay {\it pointwise} and the ground is unique if it exists.
\ei
We explain an outline of (1)-(3) above.

First we define the strongly continuous one-parameter symmetric semigroup
$\pro T$  with Kato-class potentials and general cutoff functions  by functional integral representations. Then $\KPF$ is defined by $T_t=e^{-t\KPF}$ for $t\geq 0$.
We introduce two assumptions, Assumptions \ref{ass1} and \ref{ass}, on cutoff functions of $\ms A$. The former is stronger than the later.
One advantage to define the generalized Pauli-Fierz Hamiltonian  by a path measure
 is that we need only a weak condition on cutoff functions (Assumption \ref{ass}) and external potentials.
 Then for arbitrary $\alpha\in\RR$,
Kato-class potential $V$ and cutoff function
$\hat \rho_\mu^j(x,k)$
satisfying $\hat\rho_\mu^j(x,k)\in \cb 1$, we can define $\KPF$ as a self-adjoint operator.

Secondly
we can show that
\eq{ka1}
\half(p\otimes 1+\sqrt\alpha\ms A)^2\,\dot+\,V_+\otimes 1\,\dot -\, V_-\otimes 1+1\otimes \hf
\en
is well defined
 for $V_\pm$ such that $0\leq V_+\in\loc$ and $0\leq V_-$ is relatively form bounded with respect to $p^2/2$ with a relative bound strictly smaller than one.
It is shown that
$\KPF=\kak{ka1}$
under
Assumption \ref{ass1} on cutoff functions.

Finally
it is shown that
bound states of $\KPF$ spatially exponentially  decays
{\it pointwise}.
To show the spatial exponential decay of bound states
is very important to study the properties of spectrum of Pauli-Fierz type models.
In \cite{bfs,gll,g} the spatial exponential decay of bound states is shown but
our method is completely different from them.
Since $\gr(x)=e^{tE}e^{-t\KPF}\gr$
for $\gr$ such that $\KPF\gr=E\gr$,
exponential decay of $\gr(x)$ is
 proven by means of
showing $\sup_x\|\gr(x)\|_\QQ<\infty$ and estimating  $e^{tE}\Ebb^x\lkkk e^{-\int_0^t V(B_s) ds}\rkkk$. We conclude that
\eq{m77}
\|\gr(x)\|_\QQ\leq D e^{-C|x|^{\beta}}
\en
almost everywhere $x\in\BR$,
and constants $D$ and $C$ are independent of the field mass $m$.
Here the exponent $\beta$, $\beta\geq 1$, is determined by the behavior of external potential $V$.
When $\liminf_{|x|\to\infty}V(x)<E$, we can take $\beta=1$,
and when
$V(x)=|x|^{2n}$,
$\beta=n+1$ is obtained. See Theorem \ref{14} for the detail.
Furthermore from a standard argument \cite{h9} it follows that the transformed operator $e^{i(\pi/2)N} T_te^{-i(\pi/2)N}$ is
a positivity improving semigroup, where $N$ denotes the number operator in $\QQ$. Then we  conclude that the ground state of  $\KPF$ is unique if it exists.

This paper is organized as follows:
Section 2 is devoted to constructing a strongly continuous symmetric semigroup $\pro T$ and defining
the self-adjoint operator $\KPF$.
In Section 3  we show the spatial exponential decay of bound states of $\KPF$ pointwise.
Section 4 is an appendix.

\section{Generalized Pauli-Fierz Hamiltonian}
\subsection{Definitions}
Let us begin with defining
a generalized Pauli-Fierz Hamiltonian by
a path measure.
We use the notation $\Ebb_P$ for the expectation with respect to
a probability measure $P$, i.e., $\int \cdots dP=\Ebb_P[\cdots]$.
Let $\ms S_{\rm real}=\ms S_{\rm real}(\BR)$ be the set of real-valued Schwartz test functions on $\BR$.
We set $Q=\odj \ms S_{\rm real}$.
There exist a  $\sigma$-field $\ms B$, a probability measure
$\mu$ on a measurable space
$(Q,\ms B)$
and a Gaussian random variable $\ms A(\Phi)$ indexed by $\Phi=(\Phi_1,...,\Phi_{d-1})\in
\odj L^2_{\rm real}(\BR)$
such that
\eq{21}
\mathbb E_\mu[\ms A(\Phi)]=0
\en
and the covariance
is given by
\eq{22}
\mathbb E_\mu[\ms A(\Phi)\ms A(\Psi)]=\half \jjj (\Phi_j,\Psi_j)_\LR.
\en
Throughout the scalar product on
Hilbert space $\ms L$ is denoted by $(F, G)_{\ms L}$, where it is antilinear in $F$
and linear in $G$.
 We omit $\ms L$ when no confusion arises.
For general $\Phi\in
 \od \LR$, $\ms A(\Phi)$ is defined  by
 \eq{comp}
 \ms A(\Phi)=\ms A(\Re \Phi)+i\ms A(\Im \Phi).
 \en
 Thus $\ms A(\Phi)$ is  linear in $\Phi$ over $\CC$.
 The Boson Fock space is defined by $L^2(Q,d\mu)=\QQ$. It is know that the linear hull of
\eq{23}
 \{:\ms A(\phi_1)\cdots\ms A(\phi_n):|\phi_j\in \od \LR,j=1,,.,n,n\geq 0\}
\en
is dense in $\QQ$, where $:X:$ denotes the wick product of $X$. See Section \ref{app} for the definition of Wick product.
Let us define
the free field Hamiltonian $\hf$ on $\QQ$.
Define the map $\Gamma(T):\QQ\to\QQ$ by
$\Gamma(T)1=1$ and
\eq{25}
\Gamma(T)
: \ms A(\phi_1)\cdots\ms A(\phi_n):=
:\ms A( T \phi_1)\cdots\ms A(T \phi_n):
\en
for a contraction operator $T$ on $\od \LR$.
Then $\Gamma(T)$ is also contraction on \kak{23} and can be uniquely extended to the contraction operator on the hole space $\QQ$, which is denoted by the same symbol $\Gamma (T)$.
We can check that $\Gamma(T)\Gamma(S)=\Gamma(TS)$.
Then $\{\Gamma(e^{-ith})\}_{t\in\RR}$ for a self-adjoint operator $h$ defines the strongly continuous one-parameter unitary group on $\QQ$. The self-adjoint generator of $\{\Gamma(e^{-ith})\}_{t\in\RR}$ is denoted by $d\Gamma(h)$, i.e., \eq{ka6}
\Gamma(e^{-ith})=e^{-itd\Gamma(h)},\quad t\in\RR.
\en
 Let
\eq{26}
h=\od
\omega(-i\partial),
\en
where
\eq{27}
\omega(k)=\sqrt{|k|^2+m^2},\quad m\geq0,\quad k\in\BR.
\en
Then
we set
\eq{ka8}
\hf=d\Gamma(h)
\en
 and it is called the free field Hamiltonian on $\QQ$.
Let $p=-i\partial=
(-i\partial_{x_1},...,-i\partial_{x_d})$ be momentum operators in $\LRx$.
We define the Schr\"odinger operator  $\hp$ by
\eq{51}
\hp=\half p^2+V,
\en
where $V$ denotes a real-valued external potential.
The conditions on $V$ will be required later.
The zero coupling Hamiltonian
is now given by  the self-adjoint operator
\eq{29}
\hp\otimes 1+1\otimes\hf
\en
on the Hilbert space
\eq{28}
\hhh=\LRx\otimes\QQ.
\en
The Pauli-Fierz Hamiltonian
$\PF$ is defined by replacing
$p\otimes 1$ in zero coupling Hamiltonian \kak{29} with
$p\otimes 1 +\sr  \ms A$, where
$\alpha\geq0$ is a coupling constant and
\eq{30}
\ms A_\mu =\int_\BR^\oplus \ms A_\mu(x) dx
\en
 is the so-called quantized radiation field. Here we used the identification $\hhh\cong \int_\BR^\oplus \QQ dx$.
We shall define $\ms A_\mu(x)$ below.
Let
\eq{31}
\rho_\mu^j (\cdot,x)=
(\hat \phi_\mu^j
\ov{\Psi(\cdot,x)}
/\sqrt{\omega}\check{)},\quad
j=1,..,d-1,\quad
\mu=1,...,d,
\en
where $\phi_\mu^j$ is a cutoff function and $\hat X$ (resp. $\check X$)
denotes the (resp. inverse) Fourier transform of $X$.
Note that $\hat\rho_\mu^j(k,x)=
\hat\phi_\mu^j(k)\Psi(k,x)/
\sqrt{\omega(k)}$.
Examples of cutoff functions are given letter.
The quantized radiation field is defined by
\eq{32}
\ms A_\mu(x)
=\ms A\lk
\odj  \rho_\mu^j (x)
\rk,\quad \mu=1,\cdots,d,
\en
for each $x\in\BR$.
Now we arrive at the definition of  the Pauli-Fierz Hamiltonian.
It is defined by
\eq{1}
\PF=\half(p\otimes 1+\sr  \ms A)^2+V\otimes 1+1\otimes\hf.
\en
We omit $\otimes$ for notational convenience in what follows. Then
$\PF$ is expressed as
\eq{111}
\PF=\half(p+\sr  \ms A)^2+V+\hf.
\en

\begin{assumption}
\label{ass1}
Suppose
that $\hat\rho_\mu^j\in \cb 1$ and
\eq{n1}
\omega \hat\rho_\mu^j,\
\hat\rho_\mu^j,\
\hat\rho_\mu^j/\sqrt\omega,\
\partial_{x_\mu}
\hat\rho_\mu^j,\
\partial_{x_\mu}
\hat\rho_\mu^j/\sqrt\omega\in\LRR.
\en
\end{assumption}
Under Assumption \ref{ass1}
it follows that
\begin{eqnarray}
&&\label{no1}
\|(p\cdot  \ms A+\ms A\cdot  p)F\|\leq c_1\|(p^2+\hf+1)F\|,\\
&&\label{no2}
\|\ms A\cdot\ms A F\|\leq
c_2\|(\hf+1)F\|.
\end{eqnarray}
Moreover
$\PF$ is self-adjoint on $D(p^2)\cap D(\hf)$
under Assumption \ref{ass1}.
See \cite{h11,h17,hh} for the proof.
We give examples of cutoff functions $\rho_\mu^j$.

\begin{example}{\bf (Standard Pauli-Fierz Hamiltonian)}
{\rm The standard Pauli-Fierz Hamiltonian is defined by $\PF$ with
the dimension $d=3$, $m=0$, and
$$\Psi(k,x)=e^{+ikx},\quad \hat\phi_\mu^j(k)=\hat \varphi (k) e_\mu(k,j)/\sqrt{\omega},$$
where $e(k,j)=(e_1(k,j),e_2(k,j),e_3(k,j))$, $j=1,2$,  denote  polarization vectors, and $\hat\varphi$ is an ultraviolet cutoff function.
Suppose that $\sqrt\omega\hat\varphi, \hat\varphi/\sqrt\omega,\hat\varphi/\omega\in\LR$. Then $\rho_\mu^j(k,x)\in\cb 1$ and \kak{n1} is fulfilled.
}
\end{example}

\begin{example}
{\bf(The Pauli-Fierz Hamiltonian with a variable mass)}
{\rm
 The Pauli-Fierz Hamiltonian with a variable mass $v$ instead of $m$ is studied in \cite{hid}. Then $d=3$, $m=0$,  and
$\Psi(k,x)$ is the unique  solution to
 the Lippman-Schwinger equation \cite{ike}:
\eq{lipp}
\Psi(k,x)=e^{+ikx}-\frac{1}{4\pi}\int\frac{e^{i|k||x-y|}
v(y)}{|x-y|}\Psi(k,y) dy.
\en
$\Psi(k,x)$ formally satisfies   $$(-\Delta_x+v(x))\Psi(k,x)=|k|^2\Psi(k,x),\quad k\not=0.$$
It is established that the Pauli-Fierz Hamiltonian with a variable mass has a ground state for arbitrary
values of coupling constants when
$|v(x)|\leq C (1+|x|^2)^{-\beta/2}$, $\beta>3$,  with some constant $C$. Then
 it is also seen that
\eq{a}
|\Psi(k,x)-e^{ikx}|\leq C(1+|x|^2)^{-\han}.
\en
Since
\eq{43}
\partial_{x_{\mu}}\Psi (k,x)
=ik_{\mu}e^{ikx}-
\frac{1}{4\pi}\int_{\mathbb{R}^{3}} \left(\frac{1}{|x-y|}
-{i|k|}\right)
\frac{(x_{\mu}-y_{\mu})e^{i|k||x-y|}
v(y)}{|x-y|^2}
\Psi(k,y)dy,
\en
it follows that
\eq{b}
\sup_{k\in D, x\in\RR^d_x}|\partial_{x_\mu} \Psi(k,x)|<\infty
\en
for any compact set $D$ but $D\not\ni 0$.
Let
${\rm supp} \hat\phi_\mu^j\subset D$.
Then $\rho_\mu^j\in \cb 1$ follows
from \kak{a} and \kak{b}.
In addition to condition
${\rm supp} \hat\phi_\mu^j\subset D$
 let us suppose that  $\hat\phi_\mu^j/\sqrt\omega, \sqrt\omega\hat\phi_\mu^j,
\hat\phi_\mu^j/\omega\in\LRk$,
then \kak{n1} is fulfilled.
}\end{example}

\subsection{Feynman-Kac type formulae}
Let us prepare the Euclidean version of the quantized radiation field $\ms A(\Phi)$ to construct a functional integral  representation of $e^{-t\PF}$ in the same way as \cite{h4}.
Let $Q_E=\od \ms S_{\rm real}(\BRT)$. There exist a probability measure $\mu_E$ on
a  measurable space
$(Q_E, \ms B_E)$ and a Gaussian random variable $\AA(\Phi) $ indexed by $\Phi\in \od \LRT$ such that
$$\Ebb_{\mu_E}[\AA (\Phi)]=0$$ and
the covariance is given by
$$\Ebb_{\mu_E}
[\AA(\Phi)\AA(\Psi)]=
\half \jjj(\Phi_j, \Psi_j)_{\LRT}.$$
Both  $\QQ$ and $\QQQ$ are connected through the second quantization of the family of isometry $\{\jj_t\}_{t\in\RR}$ between $\LR$ and $\LRT$:
\eq{50}
\widehat {\jj_tf}
(k_0,k)=\frac{e^{-ik_0t }}{\sqrt{\pi}}
\sqrt{\omega(k)/(\omega(k)^2+|k_0|^2)}\hat f(k).
\en
Define $\JJ_t=\Gamma(\od \jj_t):\QQ\to\QQQ$.
From the identity $\jj_t^\ast \jj_s=e^{-|t-s|\omega(-i\partial)}$ it follows that $\JJ_t^\ast \JJ_s=e^{-|t-s|\hf}$.

Set
$\ms X=C([0,\infty);\BR)$ be the set of continuous paths on $[0,\infty)$.
Let $\pro B$ denote the $d$-dimensional Brownian motion
starting at $x\in\BR$ on
$(\ms X,\ms B(\ms X),P^x)$ with the Wiener measure $P^x$. I.e.,
$P^x(B_0=x)=1$.
Let
$\cb n$ be the set of
strongly $n$-times differentiable $\LR$-valued functions on $\BR$ such that
$\sup_x\|\partial_x^z f(x)\|_\LR<\infty$ for
$|z|\leq n$.
For $f_\mu\in \cb 1$, $\mu=1,...,d$, we can define an
$\LR$-valued Stratonovich integral:
  \eq{y1}
\mmm \int_0^t f_\mu(B_s)\circ dB_s^\mu=
\int_0^t f(B_s)\cdot  dB_s+
\half \int_0^t \partial\cdot f(B_s) ds,
\en
where
$f(B_s)\cdot  dB_s=\mmm f_\mu (B_s)dB_s^\mu$ and
$\partial\cdot f(B_s)=\mmm (\partial_{x_\mu} f_\mu)(B_s)$.
 We also define
an  $\LRT$-valued Stratonovich integral
by
\eq{39}
\mmm
\int_0^t \jj_sf_\mu(B_s)\circ dB_s^\mu=
\mmm\lim_{n\to\infty}
\int_{t(j-1)/n}^{tj/n} \jj_{t(j-1)/n}
f_\mu(B_s)\circ dB_s^\mu,
\en
where $\lim_{n\to\infty}$ is a strong limit
in $L^2(\ms X;\LRT)$.
By the It\^o isometry we have the identity for $S\leq T$
\eqn
&&\Ebb^x\lkkk\lk\int_0^T j_sf(B_s)\cdot dB_s, \int_0^S j_sg(B_s)\cdot dB_s\rk_\LRT\rkkk\\
&&
\hspace{2cm}=
\mmm \int_0^S\Ebb^x\lkkk(f_\mu(B_s),g_\mu(B_s))\rkkk ds
\enn
Hence we have the bound
\eq{b1}
\Ebb^x\lkkk \left\|\mmm \int j_sf_\mu(B_s)\circ dB_s^\mu\right\|^2
\rkkk
\leq
\int_0^t ds\Ebb^x\lkkk
2\mmm
\|f_\mu(B_s)\|^2
+\half
\|\partial\cdot f(B_s)\|^2
\rkkk
\en

The next proposition is fundamental.
\bp{2}
Let $V$ be bounded. Suppose
 Assumption \ref{ass1}.
Then
\eq{3}
(F, e^{-t\PF}G)=
\int dx\Ebb^x
\lkkk
e^{-\int_0^t V(B_s) ds}
\lk
\JJ_0 F(B_0), e^{i\sr \AA (K_t)}\JJ_t G(B_t)\rk_\QQ
\rkkk,
\en
 $K_t$ is the $\od \LRT$-valued stochastic integral given by
\eq{4}
K_t=\odj\mmm
\int_0^t \jj_s\rho_\mu^j(\cdot, B_s)\circ dB_s^\mu.
\en
Here
$$\mmm\int_0^t \jj_s\rho_\mu^j(\cdot,B_s)\circ dB_s^\mu=
\int_0^t \jj_s\rho^j (\cdot,B_s)\cdot  dB_s+
\half
\int_0^t \jj_s\partial\cdot \rho^j(\cdot, B_s) ds.
$$
\ep
\proof
Suppose that
$\hat\rho_\mu^j\in \cb 2$.
Then \kak{3} is proven in the same way as
 \cite[Lemma 4.8]{h11}.
  Next we suppose that $\hat \rho_\mu^j(k,x) \in \cb 1$.
Let
$\chi\in C^\infty(\BR)$ and
 $\varphi\in C_0^\infty(\BR)$ be such that
$\chi(x)=\lkk\begin{array}{ll}
1,&|x|<1,\\
<1,&1\leq |x|\leq 2,\\
0,&2< |x|,
\end{array}
\right. $
$\varphi\geq 0$ and
 $\int \varphi(x) dx=1$. Define $\chi_N(x)=\chi(x/N)$ and $\varphi_n(x)=\varphi(x/n)n^{-d/2}$.
 Let
  \eqn
&&
\hat\rho_\mu^j(k,x)_{M,n}=
\lk
\varphi_n\ast(\rho_\mu^j(k,\cdot)\chi_N(\cdot))\rk (x),\\
&&
\hat\rho_\mu^j(k,x)_{M}=
\rho_\mu^j(k,x)\chi_M(x).
\enn
We note that
$\hat\rho_\mu^j(k,x)_{M,n}\in C_{\rm b}^\infty(\RR^d_x;\LRk)$.
Since
$\hat\rho_\mu^j(k,x)_{M,n}\to
\hat\rho_\mu^j(k,x)_{M}$ in $L^p(\RR^d_x,\LRk)$ for $1\leq p<\infty$ as $n\to \infty$,
there exists a subsequence $n'$ such that
$\hat\rho_\mu^j(k,x)_{M,n'}\to
\hat\rho_\mu^j(k,x)_{M}$ strongly in $\LRk$ for almost everywhere $x\in\BR$.
Furthermore
$\hat\rho_\mu^j(k,x)_{M}\to
\hat\rho_\mu^j(k,x)$ for each
$x\in\BR$ in $\LRk$. Then
\eq{ko2}
\lim_{M\to\infty}
\lim_{n'\to\infty}
 \hat\rho_\mu^j(k,x)_{M,n}
=
 \hat\rho_\mu^j(k,x)
 \en
  strongly in $\LRk$  for almost everywhere $x\in\BR$.
  In the same way as above we can also see that
\eq{ko3}
\lim_{M\to\infty}
\lim_{n'\to\infty}
 \partial_x^z \hat\rho_\mu^j(k,x)_{M,n}
=
 \partial_x^z \hat\rho_\mu^j(k,x)
 \en
  strongly in $\LRk$  for almost everywhere $x\in\BR$
  for $|z|\leq 1$.
Thus \kak{3} holds with $\hat\rho_\mu^j$ replaced by
$\hat\rho_\mu^j(k,x)_{M,n'}$.
$\PF$ with $\rho_\mu^j$ replaced by
$\hat\rho_\mu^j(k,x)_{M,n'}$ is
denoted by
$\PF(M,n')$.
Let $F\in C_0^\infty \otimes D(\hf)$. Then we can prove directly that
$$\lim_{M\to\infty}
\lim_{n'\to\infty}
\PF(M,n') F =\PF F.$$
Since $C_0^\infty \otimes D(\hf)$ is a core of $\PF(M,n')$ and $\PF$,
\eq{m4}
\lim_{M\to\infty}
\lim_{n'\to\infty}
e^{-t\PF(M,n')}=
e^{-t\PF}
\en
strongly.
Moreover
\eq{m2}
(F, e^{-t\PF(M,n')}G)
=\int dx\Ebb^x\lkkk
\lk \JJ_0 F(x), e^{i\sqrt\alpha \AA(K_t(M,n'))}e^{-\int_0^t V(B_s)}
\JJ_t G(B_t)\rk\rkkk,
\en
where
$K_t(M,n')$ is defined by
$K_t$ with $\rho_\mu^j(k,x)$ replaced by $\rho_\mu^j(k,x)_{M,n'}$.
 Operator $N=d\Gamma(1)$ is called
 the number operator in $\QQ$.
 Let
$F\in D(N)$. Then
the bound
$$\|\ms A(\Phi)F\|\leq 2 \|\Phi\|\|(N+1)^\han F\|$$ is known.
From \kak{m2}
and
$$|e^{i\sqrt\alpha\AA(K_t(M,n'))}
-e^{i\sqrt\alpha\AA(K_t}|\leq
|\AA(K_t(M,n')-K_t)|$$
it follows that
\begin{eqnarray*}
&&\hspace{-0.5cm}
|(F, e^{-t\PF(M,n')}G)
-(F, e^{-t\PF}G)|\\
&&\hspace{-0.5cm}
\leq
\sqrt\alpha\int dx\Ebb^x\lkkk
\lk
|\JJ_0 F(x)|,
|\AA(K_t(M,n')
-
K_t)|
e^{-\int_0^t V(B_s)}
|\JJ_t G(B_t)|
\rk\rkkk\\
&&\hspace{-0.5cm}\leq
C \sqrt\alpha
\int dx
\|(N+1)^\han F(x)\|
\Ebb^x\lkkk
\!\frac{}{}
\|K_t(M,n')
-
 K_t\|
\|G(B_t)\|
\rkkk\\
&&\hspace{-0.5cm}
\leq C \sqrt\alpha
\int dx
\|(N+1)^\han F(x)\|
\lk
\Ebb^x\lkkk
\|K_t(M,n')
-
 K_t\|^2\rkkk\rk^\han
\lk
 \Ebb^x\lkkk
\|G(B_t)\|^2\rkkk \rk^\han.
\end{eqnarray*}
We estimate
$\Ebb^x\lkkk
\|K_t(M,n')
-
 K_t\|^2\rkkk$.
 By \kak{b1} we have
 \eqn
\Ebb^x\lkkk
\|K_t(M,n')
-
 K_t\|^2\rkkk
 \leq \jjj\int_0^t
\Ebb^x\lkkk
2\mmm\|\delta\rho_\mu^j(B_s)\|^2
+
\half
\|\delta \partial\cdot \rho^j(B_s)\|^2
\rkkk ds.
\enn
where $\delta f=f-f_{M,n'}$.
By \kak{ko2} and \kak{ko3}
we see that
$$
\lim_{M\to\infty}
\lim_{n'\to\infty}
\Ebb^x\lkkk
\|K_t(M,n')
-
 K_t\|^2\rkkk=0$$
 for each $x\in\BR$.
Then  by the Lebesgue dominated convergence theorem  we have
\eq{m5}
\lim_{M\to\infty}
\lim_{n'\to\infty}
\mbox{r.h.s. }\kak{m2}=
\int dx\Ebb^x\lkkk
\lk \JJ_0 F(x), e^{i\sqrt\alpha \AA(K_t)}e^{-\int_0^t V(B_s)}
\JJ_t G(B_t)\rk\rkkk.
\en
Then \kak{3} also holds for $\rho_\mu^j\in \cb 1$. Thus the proposition follows.
 \qed

\subsection{One-parameter symmetric semigroup and generalized Pauli-Fierz Hamiltonian}
We can extend functional integral representations in Proposition \ref{2}  to more general external potentials and $\rho_\mu^j$.

\begin{definition}
{\bf (Kato class potentials)}
External potential $V: \BR \to \RR$ is called
a {Kato-class potential}
if and only if
\eq{36}
\lkk
\begin{array}{ll}
\d \sup_{x \in \BR} \int_{B_r(x)} |\la(x-y)V(y)| \, dy <\infty& d=1,\\
& \\
\d \lim_{r \to 0} \sup_{x \in \BR} \int_{B_r(x)} |\la(x-y)V(y)| \, dy = 0 &d\geq 2
\end{array}
\right.
\en
holds,
where $B_r(x)$ denotes 
the closed ball of radius $r$ centered at $x$,
and
\eq{37}
\la(x) = \left\{ %
\begin{array}{ll}
      1, &  \;\; d = 1, \\
      -\log |x|, &  \;\; d = 2, \\
      |x|^{2-d}, &  \;\;  d \geq 3.
  \end{array} \right.
\en
We denote the set of Kato-class potential by $\ms K_{kato}$.
\end{definition}
An equivalent characterization of Kato-class is as follows:
\bp{35}
A function $V$ is in $\ms {K}_{kato}$ if and only if
\eq{38}
\lim_{t \downarrow 0} \sup_{x \in \BR } \Ebb^x \left[ \int_0^t |V(B_s)| \, ds \right] = 0.
\en
\ep
\proof See e.g.,\cite{as,cfks,sim82}.
\qed
\begin{definition}
Let $\ms K$ be the set of external potential $V=V_+-V_-$ such that
$0\leq V_+\in \loc$
and
$0\leq V_-\in \ms K_{kato}$.
\end{definition}
\begin{example}
In \cite{as,sim82},
it is shown that $L_u^p(\BR)\subset \ms K_{kato}$ where
$$L_u^p(\BR)=\lkk f\left|\sup_x\int_{|x-y|\leq 1}|f(x)|^p dx<\infty\right.\rkk$$ with
\eq{pp}
p\lkk\begin{array}{ll}=1,&d=1,\\
>d/2,&d\geq2.
\end{array}
\right.
\en
In particular
let $V\in L^p(\BR)+L^\infty(\BR)$ with
\kak{pp}, then $V\in \ms K_{kato}$.
\end{example}
\begin{example}
Let $d=3$ and $\d V(x)=P(x) -\frac{a}{|x|^b}$, where $a\geq0$, $0\leq b<2$ and $P(x)=\sum_{j=0}^{2n} a_j x^j $ is a polynomial such that $a_{2n}>0$.
Then $V\in\ms K$.
\end{example}
Now we shall see that the random variable $\int_0^t V_\pm(B_s) ds$ is integrable
with respect to the Wiener measure $P^x$ for $V\in\ms K$.
\bl{n15}
Let $0\leq V\in\loc$. Then
$P^x\lk\int_0^t V(B_s) ds<\infty\rk=1$ for each $x\in\BR$.
\el
\proof
Since $V\in\loc$,
we can see that
$\Ebb^x[\int_0^t 1_NV(B_s)ds]<\infty$ for the indicator function $1_N(k)=\lkk\begin{array}{ll}1,&|k|\leq N,\\
0,&|k|>N,
\end{array}
\right.$
Then
there exists a measurable set $\ms N_N\subset \ms X$ such that $P^x(\ms N_N)=0$
and
$
\int_0^t 1_N(B_s)V(B_s)ds<\infty$ for $\omega\in \ms X\setminus \ms N_N$.
Set $\ms N=\cup_{N=1}^\infty \ms N_N$.
For $\omega\in \ms X\setminus\ms N$ we can see that
$\int_0^t 1_N(B_s(\omega))V(B_s(\omega)) ds<\infty$
for arbitary $N\geq 1$.
Let $\omega\in\ms X\setminus\ms N$.
There exists $N=N(\omega)\geq 1$ such that $\sup_{0\leq s\leq t}|B_s(\omega)|<N$.
Henceforce
$$\int_0^t V(B_s(\omega)) ds
=
\int_0^t 1_N(B_s(\omega))V(B_s(\omega)) ds<\infty,\quad \omega\in\ms X\setminus \ms N.$$
Thus
the lemma follows.
\qed

When $V_-\in\ms K_{kato}$,
it can be seen that the exponent $e^{\int_0^t V(B_s) ds}
$ is integrable with respect to $P^x$, and the supremum of $\Ebb^x\lkkk
e^{\int_0^t V(B_s) ds} \rkkk
$ in $x$ is finite. We shall check it.
\bl{n13}
Let $V\in \ms K_{Kato}$. Then there exists $\beta>0$ and $\gamma>0$ such that
\eq{561}
\sup_x\Ebb^x\lkkk
e^{\int_0^t V(B_s)} \rkkk<
\gamma e^{\beta t}
\en
Furthermore when $V\in L^p(\BR)$ with
$p\lkk\begin{array}{ll}=1,&d=1,\\
>d/2,&d\geq2,
\end{array}
\right.$ there exists $C$ such that
\eq{oto}
\beta\leq C\|V\|_p.
\en
\el
\proof
By Proposition \ref{35} there exists $t^\ast>0$ such that
$$
\alpha_t=
\sup_x\Ebb^x\lkkk
\int_0^t V(B_s) \rkkk<1
$$
 for all $t\leq t^\ast$, and
 $\alpha_t\to0$ as $t\to 0$.
It is known as Khasminskii's lemma that
\eq{41}
\sup_x\Ebb^x\lkkk
e^{\int_0^t V(B_s)} \rkkk<\frac{1}{1-\alpha_t}
 \en
 for all $t\leq t^\ast$.
By means of the Markov property of the Brownian motion we have
\eqn
\Ebb^x\lkkk
e^{-\int_0^{2t^\ast}V(B_s}
\rkkk
=
\Ebb^{x}\lkkk
e^{-\int_0^{t^\ast}V(B_s)}
\Ebb^{B_{t^\ast}}\lkkk
e^{-\int_0^{t^\ast}V(B_s)}
\rkkk
\rkkk
\leq
\lk\frac{1}{1-\alpha_{t^\ast}}\rk^2.
\enn
Repeating this procedure we can see that
\eq{562}
\sup_x\Ebb^x\lkkk
e^{\int_0^t V_-(B_s)} \rkkk
\leq \lk
\frac{1}{1-\alpha_{t^\ast}}
\rk^{[t/t^\ast]+1}
 \en
for all $t>0$, where
$[z]=\max\{w\in\mathbb Z|w\leq z\}$.
Set $\gamma =\lk
\frac{1}{1-\alpha_{t^\ast}}
\rk$ and
$\beta=
\log\lk
\frac{1}{1-\alpha_{t^\ast}}
\rk^{1/t^\ast}$.
Then \kak{561} is proven. Next we prove \kak{oto}.
Suppose $V\in L^p(\BR)$.
In the case of $d=1$
we directly see that
\eq{yu1}
\alpha_t=\int _0^ t\Ebb^x\lkkk V(B_s)\rkkk ds\leq \int_0^t (2\pi s)^{-1/2}ds\|V\|_1.
\en
Next we let $d\geq 2$ and $q$ be such that $\frac{1}{p}+\frac{1}{q}=1$.
The following estimates are due to \cite[proof of Theorem 4.5]{as}.
Let an arbitrary $\epsilon>0$ be fixed.
We have
\eqn
&&
\int_0^t
\Ebb^x\lkkk |V(B_s)|\rkkk ds\\
&&=
\int_0^t
\Ebb^x\lkkk |V(B_s)|\chi_{|B_s-x|\geq \epsilon }
\rkkk ds+
\int_0^t
\Ebb^x\lkkk |V(B_s)|\chi_{|B_s-x|< \epsilon}
\rkkk ds\\
&&\leq
t\int_{|y|\geq \epsilon}(2\pi t)^{-d/2}e^{-|y|^2/(2t)}|V(x+y)|dy+
e^t\int_0^\infty
\Ebb^x\lkkk
e^{-s}|V(B_s)|\chi_{|B_s-x|< \epsilon}
\rkkk.
\enn
It is easy to see that
\eq{242}
t\int_{|y|\geq \epsilon}(2\pi t)^{-d/2}e^{-|y|^2/(2t)}|V(x+y)|dy
\leq t (2\pi)^{-d/2}
\lk
\int e^{-q|y|^2/2}dy \rk^{1/q}
\|V\|_p.
\en
Let $f$ be the integral kernel of
$(\half p^2+1)\f$.  Then we see that
$$
\int_0^\infty ds
\Ebb^x\lkkk
e^{-s}|V(B_s)|\chi_{|B_s-x|< \epsilon}
\rkkk
\leq
\int_{|x-y|<\epsilon} f(x-y)|V(y)|dy.
$$
Since $|f(z)|\leq C \la(z)$ for
$|z|\leq\half$ with some constant $C$, we have
$$
\int_0^\infty ds
\Ebb^x\lkkk
e^{-s}|V(B_s)|\chi_{|B_s-x|< \epsilon}
\rkkk
\leq
C\int_{|x-y|<\epsilon} \la(x-y)|V(y)|dy
$$
and then
\eq{yu2}
\int_0^\infty ds
\Ebb^x\lkkk
e^{-s}|V(B_s)|\chi_{|B_s-x|< \epsilon}
\rkkk
\leq
C\lk
\int_{|z|<\epsilon} \la(z)^qdy\rk^{1/q}
\|V\|_p
\en
by the H\"older inequality.
Hence from \kak{yu1},\kak{242} and \kak{yu2},
thee exists $C_t(\epsilon)$ such that
$\alpha_t\leq C_t(\epsilon)\|V\|_p$ and
$\lim_{t\to 0}C_t(\epsilon)=
C\lk
\int_{|z|<\epsilon} \la(z)^qdy\rk^{1/q}
$.
Then for sufficiently small $T$ and $\epsilon$ we have
$\beta\leq \lk\frac{1}{1-C_T(\epsilon)\|V\|_p}\rk^{1/T}$ and
then there exists $D_T$ such that
$\beta\leq D_T\|V\|_p$.
Then \kak{oto} follows.
\qed
The functional integral representation \kak{3}  introduced in Proposition \ref{2} is well defined not only for  bounded external potentials and $\rho_\mu^j$ satisfying \kak{n1}
but also more general external potentials and $\rho_\mu^j$.
We can identify Hilbert space
$\hhh$ with $L^2(\BR\times Q)$ with the scalar product $(F,G)=\int dx(F(x), G(x))_\QQ$.
The functional integral representation of
$(F, e^{-t\PF}G)$ is
also given by
\eqn
(F, e^{-t\PF}G)=\int dx \lk
F(x), \Ebb^x\lkkk
e^{-\int_0^t V(B_s) ds}
\JJ_0^\ast e^{i\sqrt\alpha\AA(K_t)}
\JJ_t  G(B_t) \rkkk\rk_\QQ.
\enn
From this expression we shall define $\pro T$ by \kak{5} below.
\begin{assumption}
\label{ass}
We suppose that
$V\in \ms K$ and
$\hat\rho_\mu^j=\hat\rho_\mu^j(k,x)\in \cb 1$.
\end{assumption}
Note that
under Assumption \ref{ass},
$\ms A_\mu(x)$ is {\it not} relatively bounded with respect to $\hf$ in the case of $m=0$.
Under Assumption \ref{ass} however we define
the family of linear operators $\{T_t\}_{t\geq 0}$ on $\hhh$ by
\eq{5}
T_tF(x)=\Ebb^x\lkkk
e^{-\int_0^t V(B_s) ds} \JJ_0^\ast e^{i\sr \AA (K_t)}\JJ_t F(B_t)\rkkk
\en
for all $t\geq 0$.
Note that $K_t$ is well defined since
$\hat\rho_\mu^j\in \cb 1$.
\bl{6}
Suppose Assumption \ref{ass}. Then
$T_t$ is bounded on $\hhh$ for $t\geq 0$.
\el
\proof
By the definition of $T_t$ we have
$$\|T_t F\|^2_\hhh\leq \int dx \Ebb^x\lkkk e^{-2\V}\rkkk
\Ebb^x\lkkk\|F(B_t)\|^2_\QQ\rkkk.$$
Since $V\in\ms K$,
$C=\sup_x\Ebb^x\lkkk e^{-2\V}\rkkk<\infty$. Thus
$\|T_t F\|^2_\hhh\leq C
\|F\|^2_\hhh$
follows.
\qed
In what follows we shall show that $\{T_t\}_{t\geq0}$ is  a strongly continuous  one-parameter symmetric semigroup on $\hhh$.
In order to show it we introduce the second quantization of Euclidean group $\{u_t,r\}$  on $\LRT$,
where the time shift operator $u_t:\LRT\to\LRT$ is defined by
$$u_t f(x_0,{\bf x})=f(x_0-t,\bf x)$$ and
the time reflection $r:\LRT\to\LRT$ by
$$r f(x_0,{\bf x})=f(-x_0,{\bf x})$$
for $x=(x_0,{\bf x})\in\RR\times\BR$. The second quantization of $u_t$ and $r$ are denoted by $U_t:\QQQ\to\QQQ$ and $R:\QQQ\to\QQQ$, respectively.
Note that $r^\ast=r$, $rr=r^\ast r=1$, $u_t^\ast=u_{-t}$ and $u_t^\ast u_t=1$ and that
$U_t$ and $R$ are  unitary.
The time shift $u_t$, the time reflection $r$  and isometry $\jj_t:\LR\to\LRT$ satisfy the lemma below.
\bl{7}
(1)  $u_t \jj_s=\jj_{s+t}$
and $U_t \JJ_s=\JJ_{s+t}$.
(2) $r \jj_s=\jj_{-s}r$ and $RU_s=U_{-s}R$.
\el
\proof
By the definition of $\jj_s$ we have
$$\jj_s f(x)=\frac{1}{\sqrt\pi(2\pi)^{(d+1)/2}}
\int
e^{i(k_0(x_0-s)+k\cdot{\bf x})}\frac{\sqrt{\omega(k)}}{\sqrt{\omega(k)^2+|k_0|^2}}\hat f(k)
dk_0dk.$$
Then $u_t \jj_s=\jj_{s+t}$ follows, and
$U_t \JJ_s=\Gamma(u_t)\Gamma(\jj_s)=
\Gamma(u_t \jj_s)=\Gamma(\jj_{s+t})
=
\JJ_{s+t}$.
(2) is similarly proven.
\qed

\bl{52}
Suppose Assumption \ref{ass}. Then
it follows that $T_tT_s=T_{t+s}$
for all $t,s\geq 0$.
\el
\proof
By the definition of $T_t$ we have
\eq{8}
T_sT_tF(x)=\Ebb^x\lkkk e^{-\int_0^sV(B_r)dr}
\JJ_0^\ast e^{i\sr  \AA (K_s)}\JJ_s\Ebb^{B_s}
\lkkk e^{-\int_0^t V(B_r) dr}
\JJ_0^\ast e^{i\sr \AA (K_t)}\JJ_t F(B_t)\rkkk
\rkkk.
\en
Let $E_s=\JJ_s\JJ_s^\ast$, $s\in\RR$, be the family of projections.
By the formulae $\JJ_s \JJ_0^\ast=\JJ_s \JJ_s^\ast U_{-s}^\ast =E_s U_{-s}^\ast$ and $\JJ_t=U_{-s}\JJ_{t+s}$,
\kak{8} is expressed as
\begin{eqnarray*}
&&
\hspace{-0.5cm}T_sT_tF(x)\\
&&\hspace{-0.5cm}=
\Ebb^x\lkkk e^{-\int_0^sV(B_r)dr}
\JJ_0^\ast e^{i\sr  \AA (K_s)}
E_s \Ebb^{B_s}
\lkkk e^{-\int_0^t V(B_r) dr}
U_{-s}^\ast
 e^{i\sr \AA (K_t)}U_{-s}
 \JJ_{t+s} F(B_t)\rkkk
\rkkk.
\end{eqnarray*}
Since $U_s$ is unitary we have
\eq{10}
U_{-s}^\ast e^{i\sr \AA (K_t)}U_{-s}=e^{i\sr \AA (u_{-s}^\ast K_t)}
\en
as an operator, where
the exponent is given by
$$u_{-s}^\ast K_t=\odj
\mmm\int_0^t \jj_{r+s}\rho_\mu^j (B_r) \circ dB_r^\mu.$$
Let $(\fff_t)_{t\geq 0}$ be the natural filtration of the Brownian motion $(B_t)_{t\geq 0}$.
 By the Markov property
of the projections $E_t$'s \cite{sim74},
we can neglect $E_s$ in \kak{10} and we have \begin{eqnarray*}
&&T_sT_tF(x)\\
&&=
\Ebb^x\lkkk e^{-\int_0^sV(B_r)dr}
\JJ_0^\ast e^{i\sr  \AA (K_s)}
\Ebb^x
\lkkk \left.
e^{-\int_s^{s+t} V(B_r) dr}
 e^{i\sr \AA (K_s^{s+t})}
  \JJ_{t+s} F(B_{s+t})\right|\fff_s\rkkk\rkkk,
  \end{eqnarray*}
  where $\Ebb^x[\cdots|\fff_s]$ denotes the conditional expectation with respect to
  $\pro \fff$ and
$$K_s^{s+t}=\odj
\mmm \int_s^{s+t}
\jj_r \rho_\mu^j (B_r)\circ dB_r^\mu.$$
Hence we obtain that
\begin{eqnarray*}
T_sT_t F(x)
=\Ebb^x\lkkk e^{-\int_0^{s+t}
V(B_r)dr}
\JJ_0^\ast e^{i\sr  \AA (K_{s+t})}\JJ_{s+t}
  F(B_{s+t})\rkkk
  =T_{s+t}F(x)
\end{eqnarray*}
and the lemma is proven.
\qed
Next we check the symmetric property of $T_t$.
\bl{12}
Suppose Assumption \ref{ass}. Then
it follows that $T_t^\ast =T_t$ for $t\geq0$.
\el
\proof
By the functional integral representation and the unitarity of the time-reflection $R$ on $\QQQ$,
we have
\begin{eqnarray*}
(F, T_t G)
&=&
\int dx\Ebb^x\lkkk
e^{-\int_0^tV(B_s)ds}
\lk
R \JJ_0 F(B_0),
R e^{i\sr \AA (K_t)}R
R\JJ_t G(B_t)\rk
\rkkk\\
&=&\int dx\Ebb^x\lkkk
e^{-\int_0^tV(B_s)ds}
\lk
 \JJ_0 F(B_0),
 e^{i\sr \AA (r K_t)}
\JJ_{-t} G(B_t)\rk
\rkkk,\end{eqnarray*}
where the exponent is
$r K_t=\odj \mmm\int _0^t \jj_{-s}\rho_\mu^j (B_s)\circ dB_s^\mu$.
By means of the time-shift $U_{t}$ we also have
\begin{eqnarray*}
(F, T_t G)
&=&
\int dx\Ebb^x\lkkk
e^{-\int_0^tV(B_s)ds}
\lk
U_t  \JJ_0 F(B_0),
 U_t e^{i\sr \AA (r K_t)}
U_t^\ast U_t
\JJ_{-t} G(B_t)\rk
\rkkk\\
&=&
\int dx\Ebb^x\lkkk
e^{-\int_0^tV(B_s)ds}
\lk
 \JJ_t F(B_0),
  e^{i\sr \AA (u_t r K_t)}
\JJ_0 G(B_t)\rk
\rkkk,
\end{eqnarray*}
where
$u_t r K_t=\odj \mmm \int _0^t \jj_{t-s}\rho_\mu^j (B_s)\circ dB_s^\mu$.
Finally we set
$\tilde B_s=B_{t-s}-B_t$, which
equals to $B_s$ in law.
Then we have
\eq{y4}
(F, T_tG)=\int dx\Ebb^0\lkkk
 e^{-\int_0^tV(x+\tilde B_s)ds}
\lk
\JJ_t F(x),
  e^{i\sr \AA (
  \widetilde {u_t r K}_t )}
\JJ_0 G(x+\tilde B_t)
\rk
\rkkk,
\en
where
$$\widetilde {u_t r K}_t=\odj
\mmm \int_0^t  \jj_{t-s}\rho_\mu^j  (x+\tilde B_s) \circ d\tilde B_s^\mu=
\lim_{n\to\infty}
\odj  \sum_{i=1}^n
\Delta_j(i)$$
and  $\lim_{n\to\infty}$ is in the strong sense of
$L^2(\ms X;\LRT)$
and
$$
\Delta_j(i)
=
\mmm \int_{t(i-1)/n}^{ti/n}
\jj_{t-t(i-1)/n}\rho_\mu^j  (x+\tilde B_s) \circ d\tilde B_s^\mu.
$$
Then exchanging $\int dx$ and $\Ebb^0$ in \kak{y4} we have
\eqn
&&
(F,T_tG)\\
&&
=\lim_{n\to\infty}
\Ebb^0
\lkkk
\int dx
e^{-\int_0^tV(x+\tilde B_s)ds}
\lk
 \JJ_t F(x),
  e^{i\sr \AA(\odj \sum_{i=1}^n\Delta_j(i))}
\JJ_0 G(x-\tilde B_t)\rk
\rkkk
\enn
and changing variable $x-B_t$ to $x$ in $\int dx$ we have
\eqn
&&
(F,T_tG)\\
&&
=\lim_{n\to\infty}
\Ebb^0
\lkkk
\int dx
e^{-\int_0^tV(x+ B_s)ds}
\lk
 \JJ_t F(x+B_t),
  e^{i\sr \AA(\odj \sum_{i=1}^n
  \tilde \Delta_j(i))
  }
\JJ_0 G(x)\rk
\rkkk,
\enn
where
$$
\tilde \Delta_j(i)
=
-\mmm\int_{t(i-1)/n}^{ti/n}
\jj_{t-t(i-1)/n}\rho_\mu^j  (x+B_s) \circ d B_s^\mu.
$$
and
$$\lim_{n\to\infty}
\sum_{i=1}^n
\tilde
\Delta_j(i)
=
-\mmm \int_0^t \rho_\mu^j (x+B_s)\circ dB_s^\mu.$$
We thus can finally see that
$$
(F,T_t G)=
\int dx\Ebb^x\lkkk
e^{-\int_0^tV(B_s)ds}
 \lk
 \JJ_t F(B_t ),
  e^{-i\sr \AA (K_t)}
\JJ_0 G(B_0)\rk
\rkkk
=(T_tF, G).
$$
Then the lemma follows.
\qed
\bl{53}
Suppose Assumption \ref{ass}. Then
$T_t$ is strongly continuous in $t\geq 0$ on $\hhh$.
\el
\proof
Since $\|T_t\|$ is uniformly bounded and the semigroup property $T_tT_s=T_{t+s}$ is hold,
it is enough to show the weak continuity at $t=0$.
By the Lebesgue dominated convergence theorem it suffices to show that
$$\Ebb^x[(\JJ_0F(B_0),e^{i\sr\AA(K_t)}\JJ_tG(B_t)]
\to
\Ebb^x[(\JJ_0F(B_0),\JJ_0G(B_0)]
$$
as $t\to 0$ for each $x\in\BR$.
Let
\eqn
&&\Ebb^x[(\JJ_0F(B_0),e^{i\sr\AA(K_t)}\JJ_tG(B_t)]
-\Ebb^x[(\JJ_0F(B_0),\JJ_0G(B_0)]\\
&&
=\Ebb^x[(\JJ_0F(B_0),e^{i\sr\AA(K_t)}\JJ_tG(B_t)]-
\Ebb^x[(\JJ_0F(B_0),e^{i\sr\AA(K_t)}\JJ_tG(B_0)]\\
&&+
\Ebb^x[(\JJ_0F(B_0),e^{i\sr\AA(K_t)}\JJ_tG(B_0)]-
\Ebb^x[(\JJ_0F(B_0),e^{i\sr\AA(K_t)}\JJ_0G(B_0)]\\
&&+
\Ebb^x[(\JJ_0F(B_0),e^{i\sr\AA(K_t)}\JJ_0G(B_0)]-
\Ebb^x[(\JJ_0F(B_0),\JJ_0G(B_0)].
\enn
The first and second terms of the right-hand side above converge to zero as $t\to 0$, since $B_t$ and $\JJ_t$ are continuous in $t$.
We will check that the third line also goes to zero.
We have
\eqn
&&
\left|
\Ebb^x[(\JJ_0F(B_0),e^{i\sr\AA(K_t)}\JJ_0G(B_0)]-
\Ebb^x[(\JJ_0F(B_0),\JJ_0G(B_0)]\right|\\
&&\leq
\lk
\Ebb^x[\|\sr\AA(K_t)\JJ_0F(B_0)\|^2]\rk^\han
\lk\Ebb^x[\|G(B_t)\|^2]\rk^\han
\enn
We have a bound
$$
\Ebb^x\lkkk
\|\AA(K_t)\JJ_0F(B_0)\|^2
\rkkk
\leq
\|\sqrt{N+1}F(x)\|^2
\Ebb^0\lkkk
\|K_t(x)\|_\LRT^2\rkkk,
$$
where
$K_t(x)=\odj \mmm \int_0^t \jj_s\rho_\mu^j (x+B_s)\circ dB_s^\mu$.
We have
\eq{1234}
\Ebb^0\lkkk \|K_t(x)\|_\LRT^2\rkkk
\leq
\jjj
\int _0^t ds
\Ebb^x
\lkkk
2\mmm \|
\rho_\mu^j (B_s)\|^2
+\half
\|\partial\cdot \rho^j(B_s)\|^2\rkkk
.
\en
Then
$
\lim_{t\to0}
\Ebb^x\lkkk
\|\AA(K_t)\JJ_0F(B_0)\|^2\rkkk=0
$ follows and
the proof is complete.
\qed
\bt{13}
Suppose Assumption \ref{ass}.
Let $V\in\ms K$.
Then $\{T_t\}_{t\geq0}$ is a strongly  continuous  one-parameter symmetric semigroup. In particular there exists a self-adjoint operator $\KPF$ bounded below such that
\eq{y6}
e^{-t\KPF}=T_t,\quad t\geq 0,
\en
and
\eq{47}
e^{-t\KPF}F(x)=\Ebb^x\lkkk e^{-\V}\JJ_0^\ast
e^{i\sr\AA(K_t)}\JJ_t F(B_t)\rkkk.
\en
\et
\proof
This follows from
Lemmas \ref{52},\ref{12} and \ref{53}.
\qed
\begin{definition}{\bf (Generalized
 Pauli-Fierz Hamiltonians)}
Suppose Assumption~\ref{ass}.
We define a generalized Pauli-Fierz Hamiltonian with an external potential $V\in\ms K$ by a self-adjoint operator $\KPF$ in \kak{y6}.
\end{definition}
\bc{46}
Suppose Assumption \ref{ass}. Let us identify $\hhh$ with $L^2(\BR\times Q)$.
Then under this identification $e^{i(\pi/2) N}
e^{-t\KPF}e^{-i(\pi/2) N}$, $t>0$, is positivity improving. In particular the ground state of $\KPF$ is unique if it exists.
\ec
\proof
By \kak{47} we can see that
\eqn
&&(F, e^{i(\pi/2) N}
e^{-t\KPF}e^{-i(\pi/2) N}G)
\\
&&=
\int dx \Ebb^x\lkkk
\lk
\JJ_0 F(x), e^{-\V}
e^{i(\pi/2) N}
e^{i\sr\AA(K_t)}e^{-i(\pi/2) N}\JJ_t G(B_t)\rk\rkkk.
\enn
Since in \cite{h9} it is shown that
$e^{i(\pi/2) N}
e^{i\sr\AA(K_t)}e^{-i(\pi/2) N}$ is positivity improving,
$(F, e^{i(\pi/2) N}
e^{-t\KPF}e^{-i(\pi/2) N}G)
>0$ for all $0\leq F,G\in\hhh$ but $F\not=0$ and $G\not=0$.
 Then the corollary follows.
\qed
Let $L^p(\BR;\QQ)=\lkk
f:\BR\to\QQ\left|
\int \|f(x)\|^p_\QQ dx<\infty\right.\rkk$
and set the $L^p$ norm as
$\|F\|_p=(\int \|F(x)\|_\QQ^p dx)^{1/p}$.
\bc{54}
Suppose Assumption \ref{ass}. $e^{-t\KPF}$ can be extended to a bounded operator from  $L^p(\BR;\QQ)$ to itself for $1\leq p\leq\infty$.
\ec
\proof
Let
$p\not=\infty$, $p\not=1$ and
$\frac{1}{p}+\frac{1}{q}=1$.
Then we have
\eqn
\|e^{-t\KPF} F(x)\|^p_\QQ
&\leq&
\lk
\Ebb^x\lkkk
e^{-\V}\|F(B_t)\|\rkkk\rk^p\\
&\leq&
\lk
\Ebb^x\lkkk e^{-q\V}\rkkk\rk^{p/q}
\Ebb^x\lkkk \|F(B_t)\|^p_\QQ\rkkk.
\enn
Thus we have
\eqn
\int \|e^{-t\KPF} F(x)\|^p_\QQ dx\leq C\int  \|F(x)\|_\QQ^p dx.
\enn
In the case of $p=\infty$ and
$p=1$, the proof is similar.
\qed
\subsection{Quadratic form and $\KPF$}
By the functional integral representation we have
the so-called diamagnetic inequality
\eq{56}
|(F, e^{-t\PF}G)|\leq (|F|, e^{-t(\hp+\hf)}|G|)
\en
By means of the diamagnetic inequality we can see that when $|V|^\han$
is relatively bounded with respect to $(p^2/2)^\han$ with a relative bound $a\geq0$, it is also relatively bounded with respect to
$\lk
\half(p+\sqrt\alpha
\ms A)^2+\hf\rk^\han$
with a relative bound $\leq a$.
See \cite{h4}.
Let $V=V_+-V_-$ be such that
$V_+\in\loc$ and $V_-$ infinitesimally small with respect to $p^2/2$ in the sense of form.
Then under Assumption \ref{ass1} we can define the self-adjoint operator
\eq{y7}
\PF=
\half(p+\sqrt\alpha
\ms A)^2+\hf \dV
\en
by the quadratic form sum $\dot\pm$.
\bt{y8}
Let $V\in\ms K$ and suppose Assumption \ref{ass1}.
Then $\KPF=\PF$, where $\PF$ is defined by \kak{y7}.
\et
\proof
The functional integral representation of $e^{-t\PF}$ for \kak{y7}
can be given by the procedure below \cite{sim79, h4}.
Let
$$V_{n,m}(x)
=\lkk
\begin{array}{ll}
n, &V(x)\geq n.\\
V(x),&m<V(x)<n,\\
m, &V(x)\leq m.
\end{array}
\right.$$
Thus $V_{n,m}\in L^\infty(\BR)$ and then the functional integral representation of $e^{-t\PF}$ with external potential
$V_{n,m}$, which is denoted by $e^{-t\PF(n,m)}$,  is given by
Proposition \ref{35}.
By the monotone convergence theorem for forms, we can see that
$\lim_{n\to \infty}\lim_{m\to\infty}
e^{-t\PF(n,m)}=e^{-t\PF}$, where $\PF$ is defined by \kak{y7}. On the other hand the functional integral representation of
$I=(F, e^{-t\PF(n,m)}G)=\Re I+i \Im I$ is divided
into the positive part and the negative part as $$I=
(\Re I)_+-(\Re I)_- +i (\Im I)_+-i (\Im I)_-,
$$  and each term converges
as $n,m\to\infty$ by the monotone convergence theorem for integral.
Then
the functional integral representation
is given by
\begin{eqnarray}
&&(F, e^{-t\PF}G)\non\\
&&
=
\lim_{n,m\to\infty}
\int dx \Ebb \lkkk
{\lk
\JJ_0 F(B_0),
e^{-\int_0^t V_{n,+}(B_s) ds}
e^{+\int_0^t V_{m,-}(B_s) ds}
e^{i\sqrt\alpha\ms A(K_t)}
\JJ_tG(B_t)\rk}
\rkkk.\non\\
&&
\label{y9}
=
\int dx \Ebb \lkkk
{\lk
\JJ_0 F(B_0),
e^{-\int_0^t V(B_s) ds}
e^{i\sqrt\alpha\ms A(K_t)}
\JJ_tG(B_t)\rk}
\rkkk.
\end{eqnarray}
Since $V\in \ms K$, we see that
$V_+\in \loc$ and $V_-$ is infinitesimally small with respect to $p^2/2$ in the sense of form \cite[Theorem 1.12]{cfks}.
Moreover
$(F, e^{-t\KPF}G)$ equals to
the right-hand side of
\kak{y8}.
Then we conclude that $e^{-t\PF}=e^{-t\KPF}$. Thus the theorem follows.
\qed

\section{Pointwise spatial exponential decays}
In this section we show the spatial exponential decay of bound states of $\KPF$.
Let $\gr$ be  a bound state of $\KPF$ associated with eigenvalue $E$;
\eq{40}
\KPF\gr=E\gr.
\en

\begin{assumption}
\label{n14}
We say that $V=W+U\in \ms E$ if and only if $W\in \loc$, $\inf _x W(x)>-\infty$ and $0>U\in L^p(\BR)$ for some $p\lkk\begin{array}{ll}=1,&d=1,\\
>d/2,&d\geq2.
\end{array}
\right.$
\end{assumption}
Let $W+U\in \ms E$ and set $W=W_+-W_-$, where $W_\pm\geq0$ is  given by
$W_+(x)=\max\{0,W(x)\}$ and $W_-(x)=\min\{0,W(x)\}$.
Since $U\in L^p(\BR)\subset\ms K_{Kato}$, $W_-\in L^\infty\subset \ms K_{Kato}$ and $W_+\in \loc$,
we note that $\ms E\subset \ms K$.
We set
\eq{ko11}
W_\infty=\inf_xW(x).
\en
A fundamental estimate to show the spatial exponential decay of bound states is the lemma below.
\bl{carmona}
Let $V=W+U\in \ms E$.
Suppose that $\hat\rho_\mu^j\in \cb 1$.
Then for arbitrary $t,a>0$ and each $0< \alpha<\han$, there exist constants $D_1$, $D_2$ and $D_3$
such that
\eq{car}
\|\gr(x)\|_\QQ\leq
D_1
e^{D_2 \|U\|_p t}
e^{Et}
\left( D_3
 e^{-
\frac{\alpha}{4}
\frac{a^2}{t}}
e^{-tW_\infty}+ e^{-tW_a(x)}\right)\|\gr\|_\hhh,
\en
 where $W_a(x)=\inf\{W(y)| |x-y|<a\}$.
  \el
\proof
It is  a slight modification of \cite{car}.
Since $\gr=e^{tE}e^{-t\KPF}\gr$, we have
\eq{42}
\gr(x)=\Ebb^x\lkkk
\JJ_0^\ast e^{-\int_0^t V(B_s)}e^{i\sr \AA (K_t)}\JJ_t \gr(B_t)\rkkk e^{tE}.
\en
Hence for almost every $x$ it follows that
\eq{433}
\|\gr(x)\|_\QQ
\leq e^{tE}
\Ebb^x\lkkk
 e^{-\int_0^t V(B_s)}
 \|\gr(B_t)\|_\QQ
 \rkkk.
 \en
By this 
we have
$$\|\gr(x)\|_\QQ \leq e^{tE}
\lk
\Ebb^x\lkkk e^{-4\int_0^t W(B_s)ds}\rkkk\rk^{1/4}
\lk
\Ebb^x\lkkk e^{-4\int_0^t U(B_s)ds}\rkkk\rk^{1/4}
\|\gr\|_\hhh,
$$
where we used the Schwartz inequality  and 
\begin{eqnarray*}
\Ebb^x[\|\gr(B_t)\|_\QQ^2]&=&
\int (2\pi t)^{-d/2}e^{-|y|^2/{2t}}\|\gr(x+y)\|^2_\QQ dy\\
&=&
\int e^{-\pi |z|^2}
\|\gr(x+\sqrt{2\pi t}z)\|^2_\QQ dz
\leq \|\gr\|_\hhh^2.
\end{eqnarray*}
Let $A=
\{
\omega\in\ms X|
\sup_{0\leq s\leq t}
|B_s(\omega)|>a\}$.
Then it follows from a martingale inequality that
$$
\Ebb^0[1_A]\leq
2P^0(|B_t|\geq a)
=
2(2\pi )^{-d/2}
S_{d-1}
\int _{a/\sqrt t}^\infty
e^{-r^2/2}r^{d-1}
dx
\leq
\xi_\alpha
e^{-\alpha  a^2/t}
$$
with some $\xi_\alpha $ for each $0<\alpha <\han$.
Thus it follows that
\eqn
\Ebb^x\lkkk
e^{-4\int_0^t W(B_s)ds}\rkkk
&=&
\Ebb^0\lkkk 1_A e^{-4\int_0^t W(B_s+x)ds}\rkkk+
\Ebb^x\lkkk 1_{A^c}e^{-4\int_0^t W(B_s)ds}\rkkk\\
&\leq& e^{-4tW_\infty}\Ebb^0[1_A]+e^{-4tW_a(x)}\\
&\leq&
\xi_\alpha  e^{-\alpha a^2/t}e^{-4tW_\infty}+e^{-4tW_a(x)}.
\enn
Next we estimate
$\Ebb^x\lkkk e^{-4\int_0^t U(B_s)ds}\rkkk$.
Since $U$ is in Kato-class,
there exist constants $D_1$ and $D_2$ such that
$\Ebb^x\lkkk e^{-4\int_0^t U(B_s)ds}\rkkk \leq
D_1 e^{D_2\|U\|_p t}$
 by Lemmas~\ref{n13}.
Setting $D_3={\xi_\alpha}^{1/4}$,
we obtain the lemma by the inequality $(a+b)^{1/4}\leq a^{1/4}+b^{1/4}$ for $a,b\geq 0$.
\qed
For $V=W+U\in\ms E$, we define
\eq{ko8}
\Sigma=
\liminf_{|x|\to\infty}V(x).
\en
Since $U\in L^p(\BR)$, $\liminf_{|x|\to\infty}U(x)=0$ and hence
\eq{ko1-1}
\Sigma=
\liminf_{|x|\to\infty}
W(x).
\en
Moreover
$\Sigma\geq W_\infty$  holds.

\bt{17}
Suppose that
$V=W+U\in\ms E$ and $\hat\rho_\mu^j\in \cb 1$.
\bi
\item[(Confining case 1)]
Suppose that $W(x)\geq \gamma |x|^{2n}$
outside a compact set $K$
for some $n>0$ and some $\gamma >0$.
Let $0<\alpha<1/2$. Then
there exists a  constant $C_1$
such that
\eq{n20}
\|\gr(x)\|_\QQ\leq C_1
\exp\lk
-\frac{\alpha c}{16}
|x|^{n+1}\rk
\|\gr\|_\hhh,
\en
where $c=\inf_{x\in \BR\setminus K} W_{\frac{|x|}{2}}(x)/|x|^{2n}$.
\item[(Confining case 2)]
Suppose that $\lim_{|x|\to\infty}W(x)=\infty$.
Then
there exist  constants $C$ and $\delta$
such that
\eq{n20-1}
\|\gr(x)\|_\QQ\leq C
\exp\lk
-\delta |x|\rk
\|\gr\|_\hhh.
\en
\item[(Non-confining case)]
Suppose that
$
\Sigma
>E$ and $\Sigma>W_\infty$.
Let $0<\beta<1$.
Then
there exists a constant $C_2$  such that
\eq{n21}
\|\gr(x)\|_\QQ\leq
C_2
\exp\lk
-\frac{\beta}{8\sqrt 2}\frac{(\Sigma-E)}{
\sqrt{\Sigma-W_\infty}}
|x|
\rk\|\gr\|_\hhh.
\en
\ei\et
\proof
Since  $\sup_x\|\gr(x)\|_\QQ<\infty$,  it is enough to show all the statements for sufficiently large $|x|$.

(Confining case 1)
Note that
$W_{\frac{|x|}{2}}(x)\geq c|x|^{2n}$ for
$x\in \BR\setminus K$.
Then
we have bounds for
$x\in \BR\setminus K$
:
\begin{eqnarray}
|x| W_{\frac{|x|}{2}}(x)^\han
&\geq&
 c|x|^{n+1},\\
|x| W_{\frac{|x|}{2}}(x)^{-\han}
&\leq&
 c|x|^{1-n}.
 \end{eqnarray}
Inserting
$t=t(x)=W_{\frac{|x|}{2}}(x)^{-\han}|x|$ and $a=a(x)=\frac{|x|}{2}$ in \kak{car}, we have
\eq{wehave}
\|\gr(x)\|
\leq
e^{-\frac{\alpha}{16}c|x|^{n+1}}
D_1 e^{(D_2\|U\|_p+E)c|x|^{1-n}}
\lk
{D_3}
e^{c|x|^{1-n}|W_\infty|}+
e^{-(1-\frac{\alpha}{16})c|x|^{n+1}}
\rk\|\gr\|_\hhh
\en
for $x\in \BR\setminus K$.
Then  \kak{n20} follows.

(Non-confining case)
Rewrite
formula \kak{car} as
\eq{carr}
\|\gr(x)\|\leq
D_1
e^{D_2 \|U\|_p t}
\left( D_3
e^{-
\frac{\alpha}{4}
\frac{a^2}{t}}
e^{-t(W_\infty-E)}+ e^{-t(W_a(x)-E)}\right)\|\gr\|_\hhh.
\en
Then altering both $\Sigma=\liminf_{|x|\to\infty} (-W_-(x))$ and $\Sigma>W_\infty$, it is possible to choose decomposition $V=W+U\in\ms E$ such that
$\|U\|_p\leq (\Sigma-E)/2$, since $\liminf_{|x|\to\infty}U(x)=0$.
Inserting $t=t(x)=\epsilon |x|$ and $a=a(x)=\frac{|x|}{2}$ in \kak{carr}, we have
\eqn
\|\gr(x)\|
&\!\!\leq\!\!&
 D_1
e^{\|U\|_p
\epsilon |x|}
\lk {D_3}
e^{-\frac{\alpha}{16\epsilon}|x|
}
e^{-\epsilon|x|
(W_\infty -E)}
+e^{-\epsilon|x|(W_{\frac{|x|}{2}}(x)-E)}\rk
\|\gr\|_\hhh
\\
&\!\!\leq\!\!&
D_1
\lk {D_3}
e^{-\lk
\frac{\alpha}{16\epsilon}
+\epsilon
(W_\infty -E)
-\half\epsilon (\Sigma-E)
\rk |x|
}
+e^{-\epsilon
\lk
(W_{\frac{|x|}{2}}(x)-E)-\half(\Sigma-E)\rk|x|}
\rk\|\gr\|_\hhh.
\enn
Choosing
$\d \epsilon=\frac{\sqrt{\alpha/16}}{\sqrt{\Sigma-W_\infty}}$, the exponent on the first term above
turns out to
be
$$\frac{\alpha}{16\epsilon}
+\epsilon
(W_\infty -E)
-\half\epsilon (\Sigma-E)=
\half\epsilon (\Sigma-E).$$
Moreover
we see that $\liminf_{|x|\to\infty}W_{\frac{|x|}{2}}(x)=\Sigma$, and obtain
$$
\|\gr(x)\|_\QQ\leq C_2 e^{-\frac{\epsilon}{2} (\Sigma-E)|x|}\|\gr\|_\hhh$$
for sufficiently large $|x|$. Then
\kak{n21} follows.

(Confining case 2)
Finally we prove confining case 2.
In this case for arbitrary $c>0$
there exists $N$
such that
$W_{\frac{|x|}{2}}(x)\geq c$ for all $|x|>N$.
Inserting $t=t(x)=\epsilon|x|$ and $a=a(x)=\frac{|x|}{2}$ in \kak{car}, we obtain that
\eqn
\|\gr(x)\|
&\!\!\leq\!\!&
 D_1
e^{\|U\|_p
\epsilon |x|}
\lk {D_3}
e^{-\frac{\alpha}{16\epsilon}|x|
}
e^{-\epsilon|x|
(W_\infty -E)}
+e^{-\epsilon|x|(W_{\frac{|x|}{2}}(x)-E)}\rk
\|\gr\|_\hhh\\
&\!\!\leq\!\!&
 D_1
\lk {D_3}
e^{-(
\frac{\alpha}{16\epsilon}-
\epsilon \|U\|_p+\epsilon(W_\infty-E))
|x|}
+e^{-\epsilon|x|(
c-E-\|U\|_p)}\rk
\|\gr\|_\hhh
\enn
for $|x|>N$.
Choosing sufficiently large $c$ and sufficiently small $\epsilon$ such that
\eqn
&&
\frac{\alpha}{16\epsilon}-
\epsilon \|U\|_p+\epsilon(W_\infty-E)>0,\\
&&c-E-\|U\|_p>0,
\enn
we have
$\|\gr(x)\|\leq C'e^{-\delta' |x|}$ for sufficiently large $|x|$. Then \kak{n20-1}
follows.
 \qed

We give several remarks on Theorem \ref{17}.

{\bf (Independence of bose  mass $m$)}
Suppose that $\omega(k)=\sqrt{|k|^2+m^2}$.  Let $\gr$ be a normalized ground state of $\KPF$: $\|\gr\|_\hhh=1$, and $E_m=\is(\KPF)$.
It is shown that
 there exist also constants $C_1$ and $C_2$ such that
 $$\|\gr(x)\|_\QQ\leq
 C_1 e^{-C_2|x|^n},\quad n\geq1,$$ by Theorem \ref{17}.
 Since the ground state energy $E_m$ is decreasing in $m$, we can take $C_1$ and $C_2$ independent of $m<M$ with some $M$. This fact is nontrivial and useful to show the existence of ground states of the Pauli-Fierz model with $m=0$.
This is used in e.g., \cite{hid}.

{\bf (Condition $W_\infty<\Sigma$)}
When $\inf_x V(x)<\Sigma$, it is possible to decompose $V=W+U\in\ms E$ such that
$
W_\infty<\Sigma$.
In fact for arbitrary $\epsilon>0$,
there exists $y\in\BR$ such that
$$V(y)<\inf_x V(x)+\epsilon.$$
Suppose that $\inf_x V(x)+\epsilon<\Sigma$.
 Let $\ms O_y\subset \BR$ be a neighborhood of $y$.
Then define $u(x)=\lkk
\begin{array}{ll}U(x),&x\in \ms O_y,\\
0,&y\not\in \ms O_y.
\end{array}
\right.$
Let $\tilde W=W+u$ and $\tilde U=U-u$. This yields that $V=\tilde W+\tilde U\in\ms E$ and $\tilde W_\infty<\inf_x V(x)+\epsilon<\Sigma$.

{\bf (Threshold)}
The threshold is defined by $$\Sigma_\infty=
\lim_{R\to\infty}
\inf _{
F\in D_R,\|F\|=1}(F, \PF F),$$
where
$D_R=\{F\in D(\PF)|
F(x)=0, |x|<R\}$.
We note that
$\Sigma_\infty \geq\Sigma$,  and $\Sigma=\Sigma_\infty=\infty$ in confining cases.

The bound given in \cite{g} is $
\|e^{+C|\cdot|}
1_{(-\infty,\lambda]}(\PF)\|
_\hhh<\infty$,
where
$C^2+\lambda<\Sigma_\infty$.
From this the bound
\eq{g2}
\int dx
\|e^{+\delta |x|}\gr(x)\|_\QQ^2
\leq C'\|\gr\|_\hhh
\en
follows, where $$\delta<\sqrt{\Sigma_\infty-E}.$$
Theorem \ref{17}, however, gives {\it pointwise} bounds:
\eq{g1}
\|\gr(x)\|_\QQ\leq C_1\exp\lk -C_2|x|^\beta\rk\|\gr\|_\hhh,
\quad \beta\geq1.
\en
In particular
the superexponential  decay, $\|\gr(x)\|\leq C_1e^{-C_2|x|^{n+1}}\|\gr\|_\hhh$, is shown for the case of  polynomially increasing potentials
(confining case 1), while
in non-confining cases,
we show that in \kak{g1}, $\beta=1$ and
\eq{234}
C_2 <\frac{\Sigma-E}
{8\sqrt 2\sqrt{E-W_\infty}}.
\en

We give examples of external potentials.
\begin{example}
{\bf (Confining potentials)}
{\rm
Let $V=V_+-V_-$ be such that $V_+\in L_{\rm loc}^p(\BR)$ and $V_-\in L^p(\BR)$, where $p\lkk\begin{array}{ll}=1,&d=1,\\
>d/2,&d\geq2.
\end{array}
\right.$ In this case  $V\in\ms E$.
}
\end{example}

\begin{example}
{\bf (Coulomb potentials)}
{\rm
Suppose
Assumption \ref{ass1}. Then 
$$\PF=\KPF.$$
Let $V=-\alpha Z /|x|$ be the Coulomb potential.
Then $\is(\hp)=-\alpha Z/2$.
We have
$(\phi\otimes 1, \PF\phi\otimes 1)_\hhh
=(\phi, (\hp+V_{\rm eff})\phi)_\LR$
 for $\phi\in D(\half p^2)$, where
$$V_{\rm eff}(x)=\frac{\alpha}{2}
\jjj
\sum_{\mu,\nu=1}^d
(\rho_\mu^j (x), \rho_\nu^j(x))_\LR.$$
Let $V_\infty=\sup_x|
\jjj
\sum_{\mu,\nu=1}^d
(\rho_\mu^j (x), \rho_\nu^j(x))_\LR|.$
Thus
$$\is(\PF)\leq -\frac{\alpha}{2}(Z-V_\infty).$$
When $Z>V_\infty$, $\is(\PF)<\lim_{|x|\to\infty}V(x)=0$ follows for all values of coupling constant $\alpha$.
Then ground states of $\PF$ decay as $C_1e^{-C_2|x|}$ pointwise  for all values of coupling constants.
}
\end{example}

\section{Appendix}
\label{app}
In this appendix we show the unitary equivalence between $\PF$ and the Pauli-Fierz Hamiltonian defined on
$$\LR\otimes\ms F,$$
where
 $\ms F=\bigoplus_{n=0}^\infty \otimes_s^n (\oplus^{d-1}\LR)$ is the Boson Fock space over $\oplus^{d-1}\LR$.
Let $\Omega=\{1,0,0,\cdots\}\in\ms F$ be the Fock vacuum.
 The annihilation operator and the creation operator in $\ms F$
 are denoted by $\add(f)$ and $a(f)$, respectively, where
$f=(f_1,...,f_{d-1})\in \oplus^{d-1}\LR$.
They satisfy
canonical commutation relations:
\eqn
&&
[a(f), \add(g)]=\jjj(\bar f_j, g_j)_\LR,\\
&&
[\add(f),\add(g)]=0=[a(f),a(g)].
\enn
The field operator
in $\fff$ is given by
$$A(\hat \phi)=\frac{1}{\sqrt2}
(\add (\hat \phi)+a(\widetilde{\hat \phi})),$$
where
$\widetilde{\hat \phi}(k)=\hat \phi(-k)$.
  The quantized radiation field is defined by
$A_\mu=\int^\oplus_\BR A_\mu(x) dx$ under the identification
$\LR\otimes\fff\cong L^2(\BR;\fff)$ and
$
A_\mu (x)=A(\hat \rho_\mu(x))
$,
where
a cutoff function is given by
$\hat \rho_\mu(x)=
\hat \rho_\mu(k,x)=
\odj\hat \phi_\mu^j(k)
\ov{\Psi(k, x)}/\sqrt{\omega(k)}$.
Finally the free field Hamiltonian is defined by
\eq{z2}
d\Gamma(\omega)=\bigoplus_{k=0}^\infty
\sum_{i=1}^k
\underbrace{
1\otimes\cdots \stackrel{i}{\omega}\cdots\otimes 1}_{k}.
\en
Then the Pauli-Fierz Hamiltonian in $\LR\otimes\fff$ is given by
\eq{z3}
\PPF=\half(p\otimes 1+\sqrt\alpha A)^2+V\otimes 1 + 1\otimes d\Gamma(\omega).
\en
Suppose that
 $V$ is relatively bounded with respect to $\half p^2$ with a relative bound strictly smaller than one, and
that $\hat\rho_\mu^j\in \cb 1$ and
\eq{n11}
\omega \hat\rho_\mu^j,\
 \hat\rho_\mu^j,\
 \hat\rho_\mu^j/\sqrt\omega,\
\partial_{x_\mu}
\hat\rho_\mu^j,\
\partial_{x_\mu}
\hat\rho_\mu^j/\sqrt\omega\in\LRR.
\en
See Assumption \ref{ass1}.
Then $\PPF$ is self-adjoint on $D(p^2\otimes 1)\cap D(1\otimes d\Gamma(\omega))$.
Now let us see the relationship between $\QQ$ and $\fff$.
Let $\ms U:\ms F\to \QQ$ be
defined by
\eqn
&&
\ms U \Omega=1,\\
&&
\ms U:A(\hat \phi_1)\cdots A(\hat \phi_n):\Omega
=:\ms A(\phi_1)\cdots \ms A(\phi_n):,
\enn
where
the Wick product on the left hand side is defined by moving all the creation operators to the left and annihilation operators to the right without any commutation relations. While the Wick product of the left hand side is defined recursively by
$$:\ms A(\phi):=\ms A(\phi)$$
and
$$:\ms A(\phi)
\prod_{j=1}^n
\ms A(\phi_j)
:=
\ms A(\phi):\prod_{j=1}^n
\ms A(\phi_j)
:-\half \sum_{k=1}^n
(f_k,f):\prod_{j\not=k}
\ms A(\phi_j)
:.
$$
The unitary operator $\ms U$ can be extended to the unitary operator from $\fff$ to $\QQ$,
and it also implements
$$\ms U d\Gamma(\omega)\ms U^{-1}=\hf.$$
Then under \kak{n11}
it follows that
$(1\otimes \ms U)
$ maps $D(\half p^2\otimes 1)\cap D(1\otimes d\Gamma(\omega))$
to $D(\half p^2\otimes 1)\cap D(1\otimes \hf)$ and
\eq{z4}
(1\otimes \ms U)
 \PPF(1\otimes \ms U\f)=\PF.
\en

\noindent
{\bf Acknowledgments:}
 FH acknowledges support of Grant-in-Aid for
Science Research (B) 20340032 from JSPS
and
Grant-in-Aid for Challenging Exploratory Research 22654018
from JSPS.


{\footnotesize
\begin{thebibliography}{99}

\bibitem[AS82]{as}
M. Aizenman and B. Simon, Brownian motion and Harnak's inequality for Schr\"odinger operators, {\it Comm. Pure Appl. Math.} {\bf 35} (1982), 209--270.

\bibitem[BFS99]{bfs}
V. Bach,  J. Fr\"ohlich and I.M.  Sigal, Spectral analysis for
systems of atoms and molecules coupled to the quantized radiation
field, {\it Commun. Math. Phys.} {\bf 207}  (1999), 249--290.

\bibitem[BHL00]{bhl}
K. Broderix, D. Hundertmark
 and H. Leschke, Continuity properties of Schr\"odinger semigroups with
magnetic fields, {\it Rev. Math. Phys.} \textbf{12} (2000), 181--225.

\bibitem[BHLMS02]{bhlms}
V. Betz, F. Hiroshima, J. L\H{o}rinczi, R.A. Minlos and H. Spohn,
Ground state properties of the Nelson Hamiltonian --- a Gibbs
measure-based aproach, {\it Rev. Math. Phys.},  {\bf 14} (2002),
173--198.


\bibitem[Car78]{car}
R.  Carmona, Pointwise bounds for Schr\"odinger operators,
{\it Commun. Math. Phys.  } {\bf 62}  (1978), 97--106.




\bibitem[CFKS87]
{cfks} H. L. Cycon, R. G. Froese, W. Kirsch and B. Simon,
{\it Schr\"odinger operators},  Springer-Verlag Berlin-Heidelberg 1987.
\bibitem[FFG97]{ffg}
C.  Fefferman,  J.  Fr\"ohlich,  G.  M.  Graf,
Stability of ultraviolet-cutoff
quantum electrodynamics with
non-relativistic matter,  {\it Commun. Math. Phys.  } {\bf 190}  (1997), 309--330.

\bibitem[GHPS09]{ghps}  C. G\'erard, F. Hiroshima, A. Panatti and A. Suzuki, Infrared divergence of a scalar quantum field model on a pseudo Riemannian manifold, {\it Interdisciplinary Inf. Sci.}, {\bf 15} (2009), 399-421.



\bibitem[Gub06]{gub}
M. Gubinelli,
 Gibbs measures for self-interacting Wiener paths, {\it Mark. Proc. Rel. Fields }~{\bf 12} (2006), 747--766.


\bibitem[Gri01]
{g}
M. Griesemer, Exponential decay and ionization thresholds in non-relativistic quantum electrodynamics, {\it  J. Funct. Anal.}
{\bf  210}
 (2004), 321--340.

\bibitem[GLL01]
{gll}
M.  Griesemer, E.  Lieb and M.  Loss, Ground states in non-relativistic quantum electrodynamics,
{\it Invent. Math.} {\bf 145} (2001), 557--595.
\bibitem[HH08]{hh}D. Hasler and I. Herbst, On the self-adjointness and domain of Pauli-Fierz type Hamiltonians, {\it Rev. Math. Phys.} {\bf 20} (2008), 787--800.

\bibitem[Hid10]{hid}
T. Hidaka, On the existence of ground states for the Pauli-Fierz model with a variable mass,
 preprint 2010.

\bibitem[Hir97]{h4} F. Hiroshima,
Functional integral representation of a model in quantum
electrodynamics, {\it Rev. Math. Phys.} {\bf  9} (1997),
489-530.




\bibitem[Hir00-a]{h9} F. Hiroshima,
Ground states of a model in nonrelativistic quantum electrodynamics
II, {\it J. Math. Phys.} {\bf 41} (2000), 661-674.


\bibitem[Hir00-b]{h11}
\cmp{F. Hiroshima}{Essential self-adjointness of translation invariant
quantum filed models for arbitrary coupling
constants}{211}{2000}{585-613}

\bibitem[Hir01]{h17}
F.  Hiroshima,
Self-adjointness
of the Pauli-Fierz Hamiltonian for arbitrary values of coupling constants,
{ Ann. Henri  Poincar\'e}, {\bf 3} (2002), 171--201.


\bibitem[Hir07]{h26}
F. Hiroshima, Fiber Hamiltonians in nonrelativistic quantum
electrodynamics, {\it J. Funct. Anal.} {\bf 252} (2007), 314--355.

\bibitem[HIL09]{hil}
F. Hiroshima, T. Ichinose,
and
J. L\H orinczi,
Path integral representation for Schr\"odinger operator with Bernstein function of the Laplacian, preprint 2009.

\bibitem[HL08]{hl}
F. Hiroshima and J. L\H orinczi,
 Functional integral representations of the Pauli-Fierz model
with spin 1/2,
{\it J. Funct. Anal.} {\bf 254} (2008) 2127--2185.

\bibitem[Ike60]{ike}
T. Ikebe, Eigenfunction expansion asociated
with the Schroedinger operators and their applications to scattering theory, {\it Arch. Rational Mech. Anal.} {\bf 5}(1960), 1--34




\bibitem[LMS02a]{lms1}
J.  L\H{o}rinczi,
R. A.  Minlos and
H.  Spohn,
The infrared
behaviour in Nelson's model of a quantum particle coupled to a massless scalar field,
{\it  Ann. Henri  Poincar\'e} {\bf 3} (2002),
1--28.



\bibitem[Nel64]{nel}
 E. Nelson,  Schr\"odinger particles interacting with a quantized
scalar field,
 {\it Proceedings of a conference on analysis in function space},
 Ed. W. T. Martin, I. Segal, MIT Press, Cambridge 1964, p. 87.



\bibitem[Sim74]{sim74}
B.  Simon,
{\it The $P (\phi)_2$ Euclidean  (Quantum) Field Theory}, Princeton Univ. Press, 1974.


\bibitem[Sim79]{sim79}
B.  Simon,
{\it Functional Integral Representation and Quantum Physics}, Academic Press, 1979.




\bibitem[Sim82]{sim82}
B.  Simon,
 Schr\"odinger semigroups,  {\it Bull.  Amer.  Math.  Soc.  }{\bf 7}  (1982),  447--526.
{\it J. Funct. Anal.  } {\bf 32}  (1979), 97--101.

\bibitem[Spo98]{sp3}
H.  Spohn,
Ground state of quantum particle coupled to a scalar boson field,
{  Lett. Math. Phys.  } {\bf 44} (1998), 9--16.




\bibitem[Spo04]{sp4}
H. Spohn,  {\it  Dynamics of charged particles and their radiation field},
Cambridge University Press, 2004.







\end{thebibliography}
}
\end{document}